\def\rfr#1{eq. (\ref{#1})}

\def\asec{$''$ cy$^{-1}$}

\def\derp#1#2{\rp{\partial{#1}}{\partial{#2}}}
\def\dert#1#2{\frac{{{d}}{#1}}{{{d}}{#2}}}              

\def\asec{$''$ cy$^{-1}$}

\def\bar{\begin{eqnarray}}
\def\ear{\end{eqnarray}}
\def\bb{\bibitem}
\def\eqi{\begin{equation}}
\def\eqf{\end{equation}}
\def\eqia{\begin{eqnarray}}
\def\eqfa{\end{eqnarray}}
\def\rp#1#2{{#1\over#2}}

\def\lb#1{\label{#1}}

\def\oc2{$\mathcal{O}(c^{-2})$}

\def\bds#1{\boldsymbol{#1}}

\documentclass[11pt]{article}
\usepackage{amsmath,amsthm,amscd,amssymb}
\usepackage{latexsym}
\usepackage{graphicx,epsfig}
\usepackage{natbib}
\usepackage{ifpdf}
\ifpdf
\usepackage[%
  pdftitle={Instructions for use of the document class
    elsart},%
  pdfauthor={Simon Pepping},%
  pdfsubject={The preprint document class elsart},%
  pdfkeywords={instructions for use, elsart, document class},%
  pdfstartview=FitH,%
  bookmarks=true,%
  bookmarksopen=true,%
  breaklinks=true,%
  colorlinks=true,%
  linkcolor=red,anchorcolor=red,%
  citecolor=blue,filecolor=red,%
  menucolor=red,pagecolor=red,%
  urlcolor=cyan]{hyperref}
\else
\usepackage[%
  breaklinks=true,%
  colorlinks=true,%
  linkcolor=red,anchorcolor=red,%
  citecolor=blue,filecolor=red,%
  menucolor=red,pagecolor=red,%
  urlcolor=cyan]{hyperref}
\fi

\begin{document}

\noindent{\bf \LARGE{Solar System motions and the cosmological constant: a new approach}}
\\
\\
\\
{Lorenzo Iorio}\\
{\it Viale Unit$\grave{\it a}$ di Italia 68, 70125\\Bari (BA), Italy
\\tel. 0039 328 6128815
\\e-mail: lorenzo.iorio@libero.it}

\begin{abstract}
We use the corrections to  the Newton-Einstein secular precessions of the longitudes of perihelia $\dot\varpi$ of some planets (Mercury, Earth, Mars, Jupiter, Saturn) of the Solar System, phenomenologically estimated as solve-for parameters by the Russian astronomer E.V. Pitjeva in a global fit of almost one century of data with the EPM2004 ephemerides,
in order to put on the test the expression for the perihelion precession induced by an uniform cosmological constant $\Lambda$ in the framework of the Schwarzschild-de Sitter (or Kottler) space-time.  We compare such an extra-rate to the estimated corrections to the planetary perihelion precessions by taking  their ratio for different pairs of planets  instead of using one perihelion at a time for each planet separately, as done so far in literature. The answer is negative, even by further re-scaling by a factor 10 (and even 100 for Saturn) the errors in the estimated extra-precessions of the perihelia released by Pitjeva.
Our conclusions hold also for any other metric perturbation having the same dependence on the spatial coordinates, as those induced by other general relativistic cosmological scenarios and by many modified models of gravity.
Currently ongoing and planned interplanetary spacecraft-based missions should improve our knowledge of the planets' orbits allowing for more stringent constraints.
 \end{abstract}

Keywords: Experimental tests of gravitational theories; Mathematical and relativistic aspects of cosmology;  Celestial mechanics;  Orbit determination and improvement; Ephemerides, almanacs, and calendars\\

PACS: 04.80.Cc; 98.80.Jk;  95.10.Ce; 95.10.Eg; 	 95.10.Km\\

\section{Introduction}
Introduced for the first time by \citet{Ein17}
to allow static homogeneous solutions to the Einstein's equations in the presence of matter, the cosmological constant $\Lambda$, which turned out to be unnecessary after the discovery of the cosmic expansion by \citet{Hub29}, has been recently brought back mainly as the simplest way to accommodate, in the framework of general relativity, the vacuum energy  needed to explain the observed acceleration of the universe \citep{Rie98,Per99}. For the relation between the cosmological constant and the dark energy see \citep{Peb03}.   For a general overview of the cosmological constant see \citep{Car01} and references therein.   Theoretical problems concerning the cosmological constant are reviewed in \citep{Wein89}.

Since, at present, there are no other independent signs of the existence of $\Lambda$ apart from  the cosmological acceleration itself,  attempts were made in the more or less recent past to find evidence of it in phenomena occurring on local, astronomical scales with particular emphasis on the precession\footnote{For other local effects (gyroscope precession, mean motion change, geodetic precession, gravitational red-shift, deflection of light, gravitational time-delay, doppler tracking of spacecraft on escape trajectories) induced by $\Lambda$ see, e.g., \citep{Kag06,Ser06}.} of the perihelia $\omega$ of the Solar System's inner planets \citep{Isl83,Car98,Wri98,Hau03,Kra03,Dum05,Ior06,Jet06,Kag06,Ser06,Ser07,Adk07a,Adk07b}.
\section{The perihelion precession induced by a uniform cosmological constant and the confrontation with the data}
Starting from the radial acceleration \citep{Rin} \eqi \boldsymbol A_{\Lambda}=\rp{1}{3}\Lambda c^2 \boldsymbol{r},\lb{acc}\eqf where $c$ is the speed of light, imparted by an uniform cosmological constant $\Lambda$   in the framework of the spherically symmetric Schwarzschild vacuum solution with a cosmological constant, i.e. the Schwarzschild-de Sitter \citep{Stu99} or \citet{Kot18} space-time,
 \citet{Hau03} by using the Gauss equations for the variation of the Keplerian orbital elements \citep{Roy} worked out the secular, i.e. averaged over one orbital revolution, precession of the pericenter  of a test-body induced by the cosmological constant $\Lambda$   finding
\eqi\left\langle\dot\omega\right\rangle_{\Lambda}=\rp{1}{2}\left(\rp{\Lambda c^2}{n}\right)\sqrt{1-e^2}.\lb{prec}\eqf   In it
 $n=\sqrt{GM/a^3}$ is the Keplerian mean motion of the planet moving around a central body of mass $M$, $G$ is the Newtonian constant of gravitation and $a$ and $e$ are the semimajor axis and the eccentricity, respectively, of the test body's orbit.

 Here we wish to offer an alternative derivation of \rfr{prec} based on the use of the Lagrange perturbative scheme \citep{Roy}. The Lagrange equation for the pericentre is
 \eqi\dert\omega t = \rp{1}{na^2\sqrt{1-e^2}\tan i}\rp{\partial \left\langle{\mathcal{V}}_{\rm pert}\right\rangle}{\partial i} -\rp{\sqrt{1-e^2}}{na^2 e}\derp {\left\langle{\mathcal{V}}_{\rm pert}\right\rangle} e,\lb{equaz}\eqf
 where $i$ is the inclination angle to the equator of the central mass and ${\left\langle{\mathcal{V}}_{\rm pert}\right\rangle}$  is the perturbing potential ${\mathcal{V}}_{\rm pert}$ averaged over one orbital revolution.   For the Schwarzschild-de Sitter spacetime the cosmologically-induced additional potential is \citep{Hau03}
 \eqi {\mathcal{V}}_{\Lambda} = -\rp{1}{6}\Lambda c^2 r^2.\lb{dis}\eqf
 By evaluating \rfr{dis} onto the unperturbed Keplerian ellipse defined by \eqi r=a(1-e\cos E),\eqf where $E$ is the eccentric anomaly, and integrating over one orbital period $P_{\rm b}=2\pi/n$ by means of
 \eqi dt =\left(\rp{1-e\cos E}{n}\right)dE,\eqf
 \eqi\int_0^{2\pi}\left(1-e\cos E\right)^3 dE=\pi(2+3e^2)\eqf yields
 \eqi{\left\langle{\mathcal{V}}_{\Lambda}\right\rangle}=-\rp{1}{12}\Lambda c^2 a^2(2+3e^2).\lb{media}\eqf
 By inserting \rfr{media} into \rfr{equaz} one obtains just \rfr{prec}.

 \citet{Jet06}, \citet{Ser07}, \citet{Adk07a} and \citet{Adk07b} obtained, in different frameworks, the same result of \rfr{prec}. Note that $\left\langle\dot\omega\right\rangle_{\Lambda}\propto \sqrt{a^3(1-e^2)}$, where, for a uniform $\Lambda$, the proportionality factor is common to all the bodies orbiting  a given central mass. Moreover, \rfr{prec} was obtained by using the standard radial isotropic coordinate  which is commonly used in the Solar System planetary data reduction process to produce the ephemerides \citep{Est71}, so that \rfr{prec} can meaningfully be used for comparisons with the latest observational determinations   of the non-Newtonian/Einsteinian secular precessions of the longitude of the perihelia\footnote{The longitude of perihelion $\varpi$ is defined, for orbits nearly equatorial like the Solar System's ones, as the sum of the argument of perihelion $\omega$ and of the longitude of the ascending node $\Omega$; the latter one, sensitive to the out-of-plane disturbing forces, is not affected by the entirely radial $\Lambda-$induced extra acceleration.} $\varpi$ \citep{Pit05a}. Indeed, they were estimated by contrasting, in a least square sense, almost one century of data of different kinds with the suite of dynamical force models of the EPM2004 ephemerides \citep{Pit05b} which included all the standard Newtonian and Einsteinian dynamics, apart from just any exotic effects as the ones by $\Lambda$ on both the geodesic equations of motion and of the electromagnetic waves. Thus, such extra-precessions of perihelia, estimated independently of our goal, account, in principle, for any unmodelled force existing in nature.

Since the cosmological accelerated expansion yields $\Lambda\approx 10^{-56}$ cm$^{-2}$, \citet{Hau03}   concluded that the precession of \rfr{prec} is too small to be measured in the Solar System. \citet{Ior06}, \citet{Jet06}, \citet{Kag06}, \citet{Ser06}, \citet{Ser07} \citet{Adk07a} and \citet{Adk07b} used \rfr{prec} and the  extra-precessions of the inner planets of the Solar System estimated by \citet{Pit05a} to put constraints on $\Lambda$. In particular, \citet{Jet06}, after working out the effect of $\Lambda$ on the pericentre of  a general two-body system with arbitrary masses in the standard post-Newtonian gauge, used various binary pulsar systems and planets of the Solar System one at a time separately; \citet{Ser06} discussed the possibilities offered by future interplanetary ranging, especially for Pluto; \citet{Ser07} and \citet{Adk07a} obtained \rfr{prec} as a particular case in the framework of cosmological models with a non-null acceleration.  \citet{Adk07b} worked out the perihelion precessions under the action of arbitrary central forces obtaining \rfr{prec} as a particular case. In all such cases$-$and also in previous analyses performed when only data for Mercury existed\footnote{In fact, also the non-Newtonian perihelion precession of the highly eccentric ($e=0.826$) orbit of the asteroid Icarus was investigated \citep{Ic71,Ic92,Ic94a,Ic94b}, but it was never used for putting constraints on $\Lambda$. We will not use it in the present analysis.} \citep{Isl83,Car98,Wri98,Kra03}$-$the general scheme followed  was to assume the planets, or the pulsar systems considered, separately one at a time and to derive constraints on $\Lambda$ for each of them by considering it as a free parameter.
\section{Taking the ratio of the perihelia}
Here we will follow, instead, a different approach which is able to tell us something much more definite about
\rfr{acc} and \rfr{prec}.
We will construct the ratios of the estimated extra-precessions of perihelion for different pairs of planets A and B and will compare them to the corresponding ratios of the precessions of \rfr{prec} for the same planets. Thus, it is possible to construct the following quantity
\eqi \lambda_{\rm AB} = \left|\rp{\dot\varpi_{\rm A}}{\dot\varpi_{\rm B}}-\sqrt{\rp{ a_{\rm A}^3(1-e_{\rm A}^2) }{a_{\rm B}^3(1-e_{\rm B}^2)}  }\right|.\lb{la}\eqf
If \rfr{prec} is correct, independently of the value of $\Lambda$ provided that, of course, it is nonzero and small enough to assure that the perturbation approach followed to derive \rfr{prec} is appropriate, then \rfr{la} must be compatible with zero within the errors in $\dot\varpi$ and $a$ \citep{Pit05b}.    It must be noted that such an approach holds, in general, also for any other extra-acceleration term, whatever origin it may have, of the form \eqi\bds A = \mathfrak{K}\bds r,\ \mathfrak{K}\neq 0,\lb{kakka}\eqf so that our conclusions will not be restricted  to the Schwarzschild-de Sitter spacetime only.   Indeed, there are other general relativistic cosmological models \citep{Mash} and many  long-range models of modified gravity  \citep{Cog05} which are able to induce a perturbing acceleration like that of \rfr{kakka} \citep{Rug07}.
\subsection{The inner planets}
Let us start with the inner planets whose relevant orbital parameters are listed in Table \ref{tavola}.
\begin{table}
\caption{Estimated semimajor axes $a$, in AU (1 AU$=1.49597870691\times 10^{11}$ m) \citep{Pit05b}, and phenomenologically estimated corrections to the Newtonian-Einsteinian perihelion rates, in arcseconds per century (\asec), of Mercury, the Earth and Mars \citep{Pit05a}. Also the associated errors are quoted: they are in m for $a$ \citep{Pit05b} and in \asec\ for $\dot\varpi$ \citep{Pit05a}. For the semimajor axes they are the formal, statistical ones, while for the perihelia they are realistic in the sense that they
were obtained from comparison of many different
solutions with different sets of parameters and observations (Pitjeva, private communication 2005). However, the results presented in the text do not change if $\delta \dot\omega$ are re-scaled by a factor 10.}\label{tavola}

\begin{tabular}{ccccc} \noalign{\hrule height 1.5pt}
Planet & $a$ (AU) & $\delta a$ (m)  & $\dot\varpi$ (\asec) & $\delta\dot\varpi$ (\asec) \\
\hline
Mercury & 0.38709893 & 0.105 & -0.0036 & 0.0050\\
Earth & 1.00000011 & 0.146 & -0.0002 & 0.0004 \\
Mars & 1.52366231 & 0.657 & 0.0001 & 0.0005\\

\hline

\noalign{\hrule height 1.5pt}
\end{tabular}

\end{table}
The answer they give is negative. Indeed,   for the pairs A=Mars, B=Mercury and A=Earth, B=Mercury we have
\begin{equation}
\left\{
\begin{array}{lll}
\lambda_{\rm MarMer}= 7.8\pm 0.2,\\\\
\lambda_{\rm EarMer} = 4.1\pm 0.2;
\end{array}\lb{ressi}
\right.
\end{equation}
a negative result at $40-\sigma$ and $20-\sigma$ level, respectively.
 The other four pairs of inner planets yield, instead, results compatible with zero. The uncertainties in \rfr{ressi} have been conservatively worked out by propagating
 the errors in\footnote{The eccentricities are negligible.} $\dot\varpi$ and $a$ in \rfr{la} and
 linearly summing the resulting biased terms
  \eqi\delta\lambda_{\rm AB}\leq \left|\rp{\dot\varpi^{\rm A}}{\dot\varpi^{\rm B}}\right|\left(\rp{\delta\dot\varpi^{\rm A}}{|\dot\varpi^{\rm A}|} + \rp{\delta\dot\varpi^{\rm B}}{|\dot\varpi^{\rm B}|}\right) + \rp{3}{2}\left(\rp{a^{\rm A}}{a^{\rm B}}\right)^{3/2}\left(
\rp{\delta a^{\rm A}}{a^{\rm A}} + \rp{\delta a^{\rm B}}{a^{\rm B}} \right);\eqf
 the dominant sources of uncertainty are by far the perihelion rates. It is important to stress that even if $\delta\dot\varpi$ were 10 times larger than the errors released by \citet{Pit05a} and reproduced in Table \ref{tavola}$-$which are not the mere formal, statistical ones$-$we would still be forced to rule out \rfr{prec} at some $\sigma$ level.
\subsection{The outer planets}
It is interesting to use also some of the outer planets for which it was possible to estimate the corrections to the perihelion precessions  \citep{Pit06}, quoted in Table \ref{tavola2}.
\begin{table}
\caption{Estimated semimajor axes $a$, in AU (1 AU$=1.49597870691\times 10^{11}$ m) \citep{Pit05b}, and phenomenologically estimated corrections to the Newtonian-Einsteinian perihelion rates \citep{Pit06}, in arcseconds per century (\asec), of Jupiter, Saturn and Uranus. Also the associated errors are quoted: they are in m for $a$ \citep{Pit05b} and in \asec\ for $\dot\varpi$ \citep{Pit06}. For the semimajor axes they are the formal, statistical ones, while for the perihelia they are the formal ones re-scaled by a factor 10 in order to yield realistic estimates for them.}\label{tavola2}

\begin{tabular}{ccccc} \noalign{\hrule height 1.5pt}
Planet & $a$ (AU) & $\delta a$ (m)  & $\dot\varpi$ (\asec) & $\delta\dot\varpi$ (\asec) \\
\hline
Jupiter &  5.20336301 & 639 & 0.0062 & 0.036\\
Saturn & 9.53707032 & 4,222 &  -0.92 & 2.9 \\
Uranus & 19.19126393 & 38,484 & 0.57 & 13\\

\hline

\noalign{\hrule height 1.5pt}
\end{tabular}

\end{table}
The giant planets whose extra-precessions of the perihelion are at our disposal are Jupiter, Saturn and Uranus because the temporal extension of the data set used covered at least one full orbital
revolution just for such planets: indeed, the orbital periods of Neptune
and Pluto amount to about 164 and 248 years, respectively. For the external
regions of the Solar System only optical observations were used, apart
from Jupiter \citep{Pit05b}; they are, undoubtedly, of poorer accuracy with respect to
those used for the inner planets which also benefit of radar-ranging measurements,
but we will show that they are accurate enough for our purposes. Let us stress that in Table \ref{tavola2} we re-scaled by a factor 10 the otherwise formal, statistical errors in the estimated extra-rates of perihelia. The pair A=Jupiter, B=Saturn yields
\eqi\lambda_{\rm JupSat}=0.41\pm 0.06;\eqf it is incompatible with zero at about $7-\sigma$ level. If we further re-scale by 10 the error of Table \ref{tavola2} in the extra-precession of Saturn, for which Pitjeva did not use radiometric data, i.e. by 100 its formal, statistical error, we get
\eqi\lambda^{'}_{\rm JupSat}=0.4\pm 0.2,\eqf negative at $2-\sigma$ level.
In regard to Jupiter, since \citep{Pit05b} used also some radiometric data for it, we believe that a re-scaling of 10 of the formal error in its estimated perihelion extra-rate is adequate.

For the pair A=Jupiter, B=Mercury we have
\eqi\lambda_{\rm JupMer}=51\pm 12\eqf by using the figure of Table \ref{tavola} for the uncertainty in the extra-precession of Mercury; a negative result at more than $4-\sigma$ level. If, in a very pessimistic approach, we re-scale by 10 the error of Table \ref{tavola} in the Mercury's extra-rate$-$although it is not the formal one$-$we get
\eqi\lambda^{'}_{\rm JupMer}=51\pm 34,\eqf which is incompatible with zero at $1.5-\sigma$ level.   The errors in the semimajor axes, even if re-scaled by 10 or more, do not affect at all our results.  The other pairs of planets yield results compatible with zero.
%

%
%
\section{Conclusions}
In this paper we used the latest observation-based determinations of the non-Newtonian/Einsteinian precessions of the longitudes of perihelia $\dot\varpi$ of some planets of the Solar System obtained by E.V. Pitjeva (Institute of Applied Astronomy, Russian Academy of Sciences) \citep{Pit05a,Pit06} with the EPM ephemerides \citep{Pit05b} to constrain the dynamical effects induced in our planetary arena by a non-zero uniform cosmological constant in the framework of the Schwarzschild-de Sitter (or Kottler) space-time.  Such corrections to the standard perihelion precessions were determined without modelling at all the effects of $\Lambda$ on both the geodesic equations of motion of planets and electromagnetic waves carrying information on them, so that they fully account, in principle, for such effects.
The ratios of different pairs of planetary perihelia were used; by conservatively  treating the errors in the estimated extra-precessions it turns out that the expression of the $\Lambda-$induced perihelion precession is ruled out at many $\sigma$ level. It is important to note that our phenomenological approach is quite general because it holds also for any small extra-acceleration of the form $\bds A = \mathfrak{K}\bds r,\ \mathfrak{K}\neq 0:$ such a functional form is, in fact, induced not only by general relativistic cosmological models \citep{Mash} but also by  modified models of gravity (see, e.g. \citep{Cog05,Rug07}).

The present analysis relies only upon the extra-precessions estimated so far by Pitjeva; it would be useful if also other teams of astronomers would estimate independently their own corrections to the standard planetary perihelion rates by exploiting the huge records of modern observations currently available. In regard to future perspectives, our knowledge of the motion of the inner planets of the Solar System should improve in the near future thanks to the ongoing Messenger (http://messenger.jhuapl.edu/) and Venus Express (VEX) \citep{Fie08} missions to Mercury and Venus, respectively,  and to the planned BepiColombo (http://sci.esa.int/science-e/www/area/index.cfm?fareaid=30) \citep{Mil02} mission to Mercury; also Planetary Laser Ranging (PLR) \citep{Cha05}, e.g. to a lander on Mars \citep{Mer08} or to the Mercury laser altimeter \citep{Deg05}, would greatly increase the accuracy in planetary orbit determination. In regard to the outer regions of the Solar System, maybe  the processing of the radiometric data from the ongoing Saturnian mission of Cassini and from the  Jupiter's flyby of New Horizons\footnote{It took a gravity assist by Jupiter in February 2007 and should reach the orbit of Saturn in mid-2008.} (http://pluto.jhuapl.edu/), recently occurred,  might improve our knowledge of the motion of the outer planets as well in a not too far future.


\section*{Acknowledgements}
I gratefully thank E.V. Pitjeva for her information about the perihelion precessions of the outer planets of the Solar System.


\end{document}